# TAPESTRY: an interactive resource for exploring tumor-enriched splice junctions in pediatric brain tumors

Patricia J. Sullivan[1], Ryan J. Corbett[1], Chao Di[1], Alex Sickler[1], Aylar Babaei[1], Meghali Goswami[1], Jo Lynne Rokita[1,2]

[1]Center for Cancer and Immunology Research, Children's National Hospital, Washington, D.C., USA

[2]Department of Pediatrics, The George Washington University, Washington, D.C., USA

## Abstract

TAPESTRY (https://tapestry.rokitalab.com) is a database and web application for exploring recurrent tumor-enriched splice junctions (TEJs) in pediatric brain tumors, cancers that carry few protein-coding mutations and therefore few mutation-derived neoantigens. However, TEJs represent a much larger source of neoantigens as potential therapeutic targets in these tumors. TAPESTRY catalogs 45,430 recurrent TEJs across 19 histologies, identified in 1,962 RNA-seq specimens (including 1,272 independent primary tumors) from the Pediatric Brain Tumor Atlas and compared to four normal brain reference cohorts comprising adult, pediatric, developmental and cell-type-resolved tissue. Each junction is classified as tumor-specific or oncofetal; annotated for splice-event type, coding consequence, novelty relative to GENCODE v39 and overlap with Pfam and UniProt protein domains; and scored for recurrence within each histology. Users can search by gene or splicing event and filter results by histology, junction class, and annotation. Users can also generate gene-level expression heatmaps across cohorts, junction-level expression boxplots, and transcript and exon diagrams, and download all tables and figures. A documented REST API supports programmatic access, and all source code is openly available. By making tumor-selective and developmentally reactivated splicing events searchable and interpretable, TAPESTRY supports the prioritization of splice-derived targets for antibody-based, cell-based and vaccine immunotherapies in pediatric brain and central nervous system tumors.

## Introduction

Pediatric brain and central nervous system (CNS) tumors are the leading cause of cancer-related death in children, and those that resist surgery, radiation, and chemotherapy have few effective options remaining. Immunotherapies that have transformed adult oncology, including chimeric antigen receptor (CAR) T cells, antibody–drug conjugates (ADCs), and neoantigen vaccines, require targets expressed selectively by tumor cells, either on the cell surface or as intracellular peptides presented by HLA[1]. This requirement is a bottleneck in pediatric CNS tumors, which have among the lowest mutational burdens of any cancer[2], and therefore generate few of the mutation-derived neoantigens that drive responses in highly mutated adult tumors. Early clinical responses to GD2- and B7H3-directed CAR-T cells in diffuse midline glioma (DMG) show what is possible when a tumor-selective target is available, but few such targets are known[3,4].

Alternative splicing offers a large, complementary source of tumor-selective targets. The brain has one of the most complex alternative splicing programs of any tissue[5]. Splicing dysregulation is a recognized hallmark of cancer[6], as tumors can skip exons, retain introns, and activate cryptic splice sites, producing transcript and protein isoforms not found in normal cells. Junctions absent from normal tissue and the peptides and proteoforms they encode are therefore attractive candidates for tumor-selective therapy[7]. The difficulty lies in finding these events among the many that each tumor produces. For example, rMATs-turbo[8] reports a median of ~130,000 splicing events per tumor in the Pediatric Brain Tumor Atlas (PBTA) cohort, of which only a small fraction are tumor-enriched, recurrent across patients, and located in functionally relevant protein regions. Isolating that fraction requires comparing each junction against normal references that include the developing brain, annotating its transcript and protein context, and scoring its recurrence within histologies.

Existing resources do not fill this gap for pediatric CNS tumors. Splicing quantification tools such as rMATS[8] produce event calls, not a curated catalog of tumor-selective candidates. Normal-tissue references such as GTEx[9] provide adult baselines but lack the fetal and developmental brain expression data[10] needed to distinguish a tumor-specific junction from a reactivated developmental, or oncofetal, one. Existing pan-cancer splicing databases and atlases are built largely on adult cohorts and are not organized around splice junctions as prioritizable therapeutic targets. As a result, the tumor-enriched splicing signal already present in harmonized pediatric cohorts such as the Open Pediatric Cancer Project (OpenPedCan)[11], built from Children's Brain Tumor Network specimens[12], remains difficult to mine.

Targets already identified from these cohorts show that this signal is therapeutically actionable. Recurrent *NRCAM* microexon skipping generates a tumor-specific cell-surface proteoform in DMG that is amenable to antibody-based targeting[13]. *CLK1* exon 4 inclusion, a developmental splicing event normally suppressed after fetal development, persists across PBTA histologies, predicts poor event-free survival, and can be targeted with small-molecule *CLK* inhibitors or splice-modulating oligonucleotides[14]. Splice-derived antigens identified through a related discovery workflow are now being evaluated in a Phase 1 trial of personalized splice-derived tumor-specific antigen-directed T cells for children with brain tumors (NCT06193759).

Here we present TAPESTRY (Tumor Alternative PEdiatric Splicing visualizaTion and queRYing; https://tapestry.rokitalab.com), a database of recurrent TEJs in pediatric brain tumors. TAPESTRY integrates splice-junction quantifications from 1,272 independent primary PBTA tumor RNA-seq specimens with four normal brain reference cohorts, classifies each junction as tumor-specific or oncofetal, and annotates it with transcript, coding and protein-domain context. The TEJ discovery approach is implemented in the tumor-enriched splicing pipeline [15], and the method and its biological findings will be reported separately. Here, we describe the database content, the web interface and REST API through which it can be searched, visualized and downloaded, and use cases illustrating how TAPESTRY supports the prioritization of splice-derived therapeutic targets.

## Materials and methods

### *Data sources and cohort*

TAPESTRY is built on RNA-sequencing from the Kids First Pediatric Brain Tumor Atlas (PBTA; dbGaP phs002517), with clinical annotations and expression context harmonized through the OpenPedCan project[11]. The current data release comprises 1,880 tumor RNA-seq specimens, of which 1,272 independent primary tumors are used for recurrence (Section 2.7), spanning 19 tumor histologies. TAPESTRY also includes 82 patient-derived cell-line models from the Children's Brain Tumor Network for validation of catalog-nominated junctions in tractable preclinical systems. Tumor junctions were compared against four independent normal/control reference cohorts that together span the postnatal-to-developmental axis of the human brain (**Table 1**).

***Table 1.*** *Reference tissue and cell of origin cohorts used for calling tumor-enriched junctions.*

| Reference cohort | Source | N | Role in the pipeline |
|---|---|---|---|
| GTEx brain (ages 20–39, per region)[9] | GTEx (dbGaP phs000424) | 113 | Adult postnatal normal brain |
| Evo-Devo (Forebrain, Hindbrain; fetal vs postnatal)[16] | Kaessmann developmental atlas (ENA PRJEB26969) | 106 | Developmental brain; **the fetal split defines the oncofetal call** |
| Normal pediatric brain (per primary site) | GSE243682 | 7 | Postnatal pediatric normal brain |
| Pediatric brain cell types[17] | GSE73721 | 15 | Cell-type-resolved postnatal brain (used for annotation/visualization, not the enrichment test) |

All data are referenced to genome build **GRCh38/hg38** with **GENCODE v39** gene annotation. *The Evo-Devo atlas is single-end RNA-seq, which has reduced splice-junction detection sensitivity relative to paired-end data.*

*Splice-junction quantification*

RNA-seq reads were aligned with STAR v2.7 to GRCh38 (GENCODE v39), following the Kids First RNA-seq harmonization workflow[11], and splice junctions were quantified from the resulting alignments with rMATS-turbo v4.3[8]. Four alternative-splicing event classes were identified — skipped exon (SE), retained intron (RI), and alternative 3′ and 5′ splice sites (A3SS, A5SS). For each junction in each sample, junction expression is normalized to counts per million (CPM):

$$junction\ CPM = \frac{junction\ count}{used\ read\ count} \times 10^6$$

where `used read count` is the rMATS input read count; junctions require at least three supporting reads at a boundary to be retained. TAPESTRY stores this per-sample CPM together with the underlying junction read count and the host gene's expression (gene TPM).

*Construction of the TEJ catalog*

The TEJ catalog loaded into TAPESTRY was generated by the tumor-enriched splicing pipeline (https://github.com/rokitalab/pbta-tumor-enriched-junctions; [15]). Briefly, junction read counts were normalized to CPM and corrected for RNA library preparation type. For each junction in each tumor sample, two enrichment statistics were computed against the postnatal reference cohorts (GTEx brain, postnatal Evo-Devo brain and normal pediatric brain). The fold-change (FC) is the tumor CPM divided by the highest mean CPM among the postnatal reference groups. The signal-to-noise ratio (SNR) is the difference between the tumor CPM and the reference mean, divided by the reference standard deviation, taking the lowest value across reference groups. A junction was called tumor-enriched in a sample when FC and SNR both exceeded 5 and its highest mean CPM in any postnatal reference group was below 10. Junctions whose enrichment was explained by host-gene overexpression rather than a change in percent spliced in (PSI) were removed, and each remaining junction was assigned a splicing class (**Table 2**). Junctions expressed more highly in prenatal than in postnatal Evo-Devo brain (FC > 2 and SNR > 2) were labeled oncofetal; all others were labeled tumor-specific. Junctions were retained only if recurrent in at least one histology, defined as detection in at least 3 independent primary tumors. FC, SNR and maximum mean CPM are stored for every junction, computed against both the postnatal references and all references. The Explore interface exposes them as adjustable filters, so users can apply thresholds stricter or more permissive than those used to build the catalog.

***Table 2.*** *Abbreviations for each type of splicing event.*

| Code | Splicing Event |
|---|---|
| EI / ES | Exon inclusion / exon skipping |
| RI | Retained intron |
| A3+ / A3− | Alternative 3′ splice site (long / short) |
| A5+ / A5− | Alternative 5′ splice site (long / short) |

*Junction annotation*

Each junction was annotated against the GENCODE v39 reference (genome build GRCh38/hg38):

- **Status** — a three-way classification against annotated exon boundaries: Annotated junction (both splice sites match known boundaries), Novel splice site (one boundary known, one novel), or Novel junction (the exon–exon junction is supported by no annotated transcript).
- **Coding context** — whether the junction falls within a coding sequence (`in_cds`) and its predicted consequence.
- **Protein domains** — junctions that are mapped to overlapping protein features from two sources: Pfam domains (via the annoFuseData Bioconductor data package[18], joined by gene symbol, with overlap classified as complete / partial / none) and UniProt topological domains (via `bedtools intersect` against UCSC hg38 topology tracks, restricted to CDS regions of protein-coding transcripts). Domain overlap is what connects a junction to a potentially targetable protein feature.

*Software architecture and implementation*

TAPESTRY is a three-tier open-source application (**Figure 1**). The **Data Pipeline** (https://github.com/rokitalab/TAPESTRY-data-preprocessing) is a Python/pandas ETL (psycopg 3) that ingests harmonized histology tables, the annotated recurrent-junction table, per-sample CPM matrices (RDS/qs2, read via Rscript), and protein-domain annotations, and loads them into a normalized PostgreSQL schema (`patient`,

`sample`, `tej`, `tej_domain`, `sample_tej`, `tej_cpm`, `tej_recurrence`). Materialized summary views (`tej_gene_summary`, `tej_histology_summary`, `gene_junction_summary`) precompute the per-gene and per-histology aggregates the API serves. Input data are versioned and distributed via Amazon S3 with md5 manifests; the pipeline is containerized (`docker compose`) for one-command local reproduction.

The **API** (https://github.com/rokitalab/TAPESTRY-API) is a Flask (≥3.0) / flask-restx (≥1.2) service served by gunicorn (Python 3.11, 2 workers, port 8080), with Flask-Caching and CORS. All filtering and sorting are done in SQL; the API contains no summarization logic of its own, reading directly from the materialized summary views. All endpoints use GET and require no authentication. Interactive OpenAPI/Swagger documentation is served at `/tapestry-api/doc/`. The database is PostgreSQL (Amazon RDS in the production deployment).

The **Frontend** (https://github.com/rokitalab/TAPESTRY) is a React application (Material UI, D3) served via nginx and containerized with Docker. It communicates with the API at `/tapestry-api`.

*Interactive features*

- **Home** — quick search by gene symbol or histology, with cohort-level summary statistics.
- **Explore** — the primary interface: a filterable, sortable, paginated table of TEJs. Server-side filtering selects a histology; additional client-side filters cover gene, junction status, splice-event type, consensus specificity, cohort scope (all vs postnatal), and sliders for minimum fold-change, minimum SNR, and maximum mean CPM.
- **Gene-level views** — for a selected gene: the event table, exon-block diagrams, transcript visualizations with GENCODE biotype toggles, and overlaid Pfam/UniProt protein domains. Plots include hyperlinks to PedCbioPortal for PBTA samples.
- **Event-level views** — for a selected junction: tumor-enrichment statistics against the cohort and a cross-cohort **junction-expression heatmap** showing per-sample CPM across tumor histologies, the four normal/control cohorts, patient-derived cell lines, and evo-devo developmental timepoints — directly supporting wet-lab target triage (e.g. confirming a junction's presence in a model cell line and its absence from normal and developmental brain).

- **Export** — tables export to Excel; all plots export to publication-quality SVG, PDF, and TIFF.
- **Docs / About** — a glossary of terms, use-case walkthroughs, and full data provenance.

## Results

The current release of TAPESTRY (v1.0.0, September 2026) contains 1,880 PBTA tumor specimens from 19 histologies, 241 reference samples from four normal brain cohorts and 82 patient-derived cell lines, together with a catalog of 45,430 recurrent TEJs in 10,095 genes (**Figure 2**). Of these TEJs, 3,928 (8.6%) are oncofetal (**Figure 2B**). Exon inclusion is the most frequent splicing class (21,198; 46.7%), followed by exon skipping (6,824; 15.0%) (**Figure 2C**). The genes with the most TEJs are *GFAP* (90), *LRP1* (82) and *COL20A1* (75) (**Figure 2D**). Specimen numbers vary widely across histologies, from 425 low-grade gliomas to 12 pineoblastomas (**Figure 2A**). Summary statistics are recomputed with each release and are available on the About page and through the API.

The **Explore** interface narrows the catalog to a short list of candidates. Users can filter by histology, gene, junction status, splicing class, and oncofetal status, and set thresholds on fold-change, SNR, and maximum mean CPM in normal brain. For example, filtering to oncofetal exon-skipping junctions in Other high-grade gliomas, with fold-change ≥ 10, SNR ≥ 10 and maximum mean CPM ≤ 1 across all reference cohorts, returns 115 candidates ranked by enrichment statistics (**Figure 3A**). One of these, a *HSPG2-201* exon5:exon3 skipping junction, is expressed in early fetal forebrain and hindbrain with lower expression at every later prenatal and postnatal stage sampled (**Figure 3B**), typical for an oncofetal event. The skipped exon is absent from the canonical transcript and does not overlap an annotated protein-coding domain (**Figure 3C**).

*NRCAM* microexon skipping has already been reported as a tumor-specific, antibody-targetable cell-surface proteoform in high-grade glioma[13], making it a useful test of whether TAPESTRY recovers a validated target. We therefore checked whether TAPESTRY recovers this event. Two *NRCAM-202* exon-skipping junctions, exon6:exon4 (Δex5) and exon19:exon17 (Δex18), are in the catalog as tumor-enriched, non-oncofetal junctions, and each is detected in more than 150 tumor samples (**Figure 4A**). The exon6:exon4 junction is expressed across primary tumor histologies, most highly in low-grade glioma, and is low or undetectable in developmental, pediatric and adult GTEx brain (**Figure 4B**). However, it is also expressed in the astrocyte samples of the pediatric brain cell-type cohort. Taken together with its high expression in glial tumors, this suggests that the junction reflects an astrocytic splicing program that is enriched in tumors, rather than being absent from all normal cells. This example illustrates how cell-type-resolved reference flags junctions that bulk normal tissue would miss. Across the *NRCAM* locus, both junctions are expressed in the same tumor samples (**Figure 4C**).

**Discussion**

TAPESTRY makes the tumor-enriched splicing signal in pediatric CNS tumors searchable and interpretable. It differs from existing splicing resources in three ways. First, it is built on pediatric CNS tumors and pairs every junction with its expression across adult, pediatric, developmental and cell-type-resolved brain, so users can see where else a junction is expressed rather than relying on a binary call. Second, its enrichment thresholds are exposed as adjustable filters rather than fixed cutoffs. Third, each junction is linked to its recurrence, protein-domain context and expression in patient-derived cell lines, the information needed to decide whether a candidate is worth validating. The oncofetal label matters for safety assessment: an oncofetal junction may still be a viable target in a patient past the developmental window, but it carries a different risk profile from one absent from all normal brain.

The resource has limitations. Junction calls depend on rMATS-turbo and on the depth and composition of the normal reference cohorts, some of which (pediatric normal brain, n = 7; pediatric brain cell types, n = 15) are small; enrichment thresholds are heuristics rather than a calibrated statistical model, which is why TAPESTRY exposes them as adjustable filters; junction-level evidence does not by itself establish that a stable protein isoform or an HLA-presented peptide is produced; and the current catalog is restricted to junctions recurrent within a histology, by design excluding private events. Experimental validation — at the transcript, protein, and immunopeptidome levels — remains essential, and TAPESTRY is intended to prioritize candidates for exactly that validation.

A priority next step is to deepen the developmental and cross-tissue context of the normal references. The current fetal reference (Evo-Devo) is single-end and covers only two brain regions at low sample size, which limits sensitivity across the neonatal-to-adolescent window and provides no non-CNS baseline for assessing off-target risk. The developmental GTEx (dGTEx) resource[19] — paired-end RNA-seq spanning neonatal through adolescent donors across dozens of tissues — together with adult GTEx across all tissues would let TAPESTRY stage each junction along a developmental trajectory and score it for CNS-restriction, refining the oncofetal classification and flagging junctions expressed in critical non-CNS tissues before they are nominated as systemic targets. Further directions include incorporating additional pediatric tumor types as harmonized data become available, adding immunopeptidomics and predicted-neoepitope layers, integrating cell-line and PDX model availability directly into target views, and linking candidate junctions to designed ASO or CAR-T reagents. By releasing the full pipeline, database schema, API, and frontend as open source with versioned data, we intend TAPESTRY to be both a directly usable resource and a template for tumor-enriched splicing atlases in other cancers.

## Use of AI tools

During the preparation of this work, the authors used Claude (Anthropic) to assist with software development of the TAPESTRY web application, API and data pipeline and with editing, condensing and restructuring the manuscript text. AI-assisted code was reviewed and tested by the authors through code review and verification of API outputs against the underlying database. All manuscript text was reviewed and revised by the authors, and all scientific content, analyses and conclusions are the authors' own. The authors take full responsibility for the content of this publication.

## Data Availability

Web application: https://tapestry.rokitalab.com

Frontend source: https://github.com/rokitalab/TAPESTRY

Data pipeline source: https://github.com/rokitalab/TAPESTRY-data-preprocessing[20]

Upstream tumor-enriched splicing method: https://github.com/rokitalab/pbta-tumor-enriched-junctions[21]

Data: PBTA RNA-seq data are available under controlled access through dbGaP (phs002517) and the Kids First Data Resource Portal. The TEJ catalog can be downloaded through the API. Source code is released under a BSD-3 license.

**Acknowledgements**

We thank the patients and families who contributed tumor samples and clinical data to the Children's Brain Tumor Network (CBTN) and the Pacific Pediatric Neuro-Oncology Consortium (PNOC); this work would not be possible without their generosity. We thank CBTN, PNOC and their member institutions, investigators and staff for specimen collection, clinical annotation and data sharing, and the Gabriella Miller Kids First Data Resource Center for data harmonization and access. We also thank the generators of the GTEx, Evo-Devo developmental atlas, the normal pediatric brain (GSE243682) and brain cell-type (GSE73721) datasets, and the Cancer Dependency Map for making their data accessible and/or publicly available.

**Funding**

This work was supported by the anonymous investors who support the Brain Tumor Institute at Children's National Hospital (JLR) and the Children’s Cancer Foundation (RJC), in part.

**Author Contributions**

Conceptualization: PJS, RJC, AN, JLR

Methodology: PJS, RJC, CD, JLR

Software: PJS, CD

Validation: PJS, RJC, MG, JLR

Formal Analysis: RJC, PJS, AS, AB

Investigation: PJS, RJC, JLR, MG

Resources: JLR

Data Curation: JLR, AS, CD, RJC

Writing - Original Draft: PJS, JLR

Writing - Review & Editing: PJS, RJC, JLR

Visualization: PJS

Supervision: JLR

Project Administration: JLR

Funding Acquisition: JLR, RJC

**Competing interests**

The authors declare no competing interests.

**Figure Legends**

**Figure 1 – TEJ discovery pipeline.** Tumor-enriched junctions (TEJs) are called from 1,916 PBTA tumor specimens (19 histologies) against four normal reference cohorts (241 samples in total: GTEx, Evo-Devo, normal pediatric brain and pediatric brain cell types). Junctions enriched in tumors relative to postnatal normal brain, and therefore reflecting differential splicing rather than host-gene up-regulation, are called tumor-enriched. Those expressed more highly in prenatal than postnatal brain are labeled oncofetal. Only junctions recurrent within a histology among the 1,272 independent primary tumors are retained, giving a catalog of 45,430 recurrent TEJs. The catalog is loaded into a PostgreSQL database, served through a REST API and displayed in the TAPESTRY web application.

**Figure 2 — Cohort and catalog summary.** Summary values correspond to the active data release and are accessible dynamically through REST API endpoints. **(A)** Breakdown of sample counts across tumor histologies. **(B)** Oncofetal proportion of TEJs. **(C)** Representation of TEJs categorized by splicing class (exon skipping, exon inclusion, retained intron, and alternative 5′/3′ boundaries). **(D)** Top genes harboring the most TEJ events cohort-wide.

**Figure 3 – Explore filtering and an oncofetal example. (A)** The Explore interface filtered to oncofetal, exon-skipping junctions in other high-grade glioma (fold-change ≥ 10, SNR ≥ 10, max mean CPM ≤ 1), returning 115 candidate junctions ranked by tumor-enrichment statistics. **(B)** Developmental expression of the highlighted *HSPG2-201* exon5:exon3 junction across the Evo-Devo forebrain and hindbrain time course. Small points are samples; large points are the mean per timepoint. **(C)** Transcript view of the junction. The reference transcript *HSPG2-203* is shown with its UniProt domains. In the junction transcript *HSPG2-201*, the red arc marks the junction, and the red exon is the skipped exon 4.

**Figure 4 – NRCAM microexon skipping. (A)** TAPESTRY event table for two linked, tumor-specific *NRCAM* exon-skipping junctions (*NRCAM-202* exon6:exon4 and exon19:exon17) with tumor-enrichment statistics. **(B)** Per-sample expression of the exon6:exon4 junction across primary tumor histologies, cell-of-origin, developmental, pediatric, and adult (GTEx) reference cohorts. **(C)** Cross-cohort junction-expression heatmap across the *NRCAM* locus, highlighting the two linked microexon skipping junctions (Δex18, Δex5) against the gene and transcript model.

Figure 1

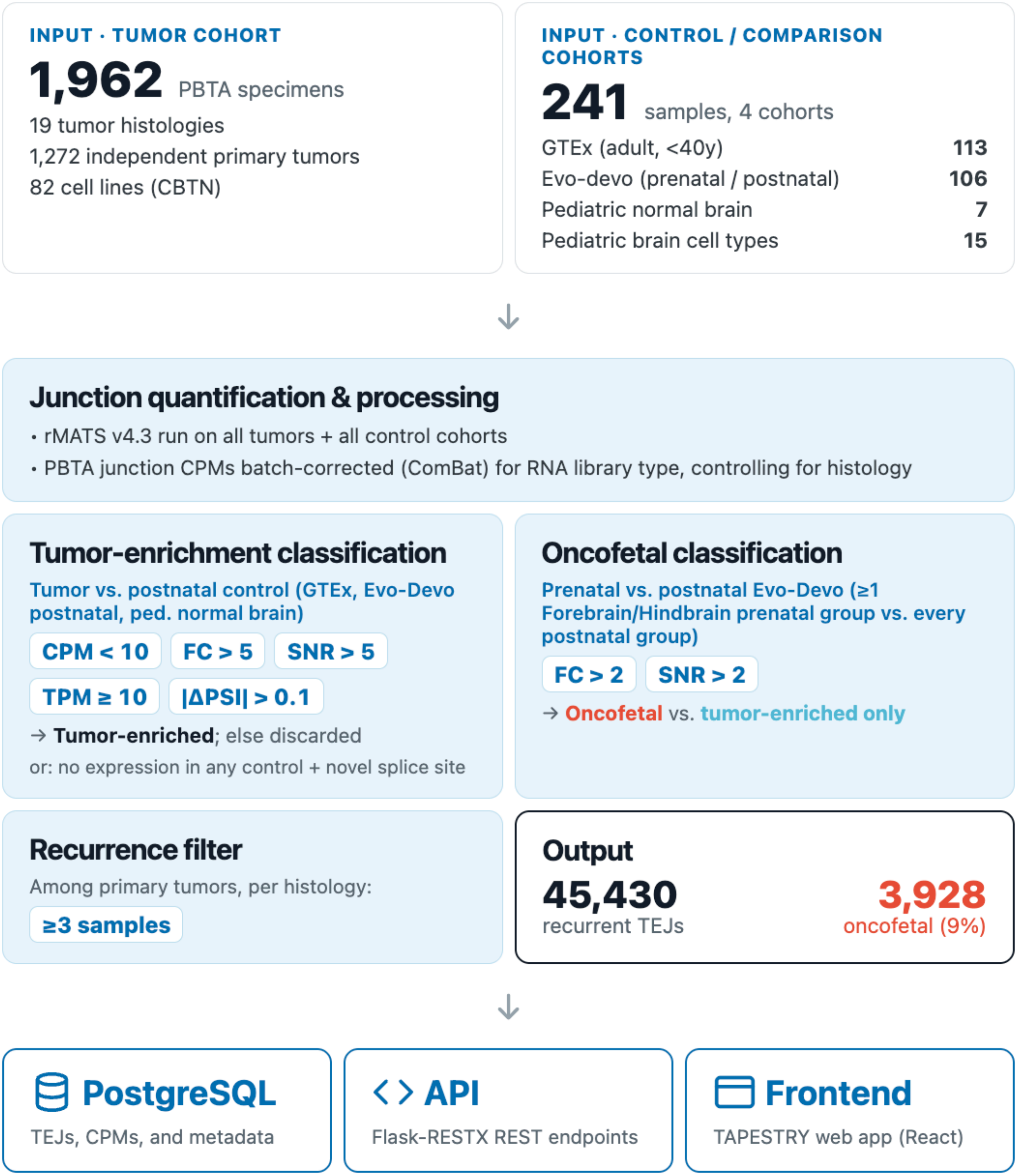
INPUT · TUMOR COHORT
1,962 PBTA specimens
19 tumor histologies
1,272 independent primary tumors
82 cell lines (CBTN)
INPUT · CONTROL / COMPARISON COHORTS
241 samples, 4 cohorts
GTEx (adult, <40y) 113
Evo-devo (prenatal / postnatal) 106
Pediatric normal brain 7
Pediatric brain cell types 15
Junction quantification & processing
• rMATS v4.3 run on all tumors + all control cohorts
• PBTA junction CPMs batch-corrected (ComBat) for RNA library type, controlling for histology
Tumor-enrichment classification
Tumor vs. postnatal control (GTEx, Evo-Devo postnatal, ped. normal brain)
CPM < 10
FC > 5
SNR > 5
TPM ≥ 10
|ΔPSI| > 0.1
→ Tumor-enriched; else discarded
or: no expression in any control + novel splice site
Oncofetal classification
Prenatal vs. postnatal Evo-Devo (≥1 Forebrain/Hindbrain prenatal group vs. every postnatal group)
FC > 2
SNR > 2
→ Oncofetal vs. tumor-enriched only
Recurrence filter
Among primary tumors, per histology:
≥3 samples
Output
45,430 recurrent TEJs
3,928 oncofetal (9%)
PostgreSQL
TEJs, CPMs, and metadata
API
Flask-RESTX REST endpoints
Frontend
TAPESTRY web app (React)

**Figure 2**

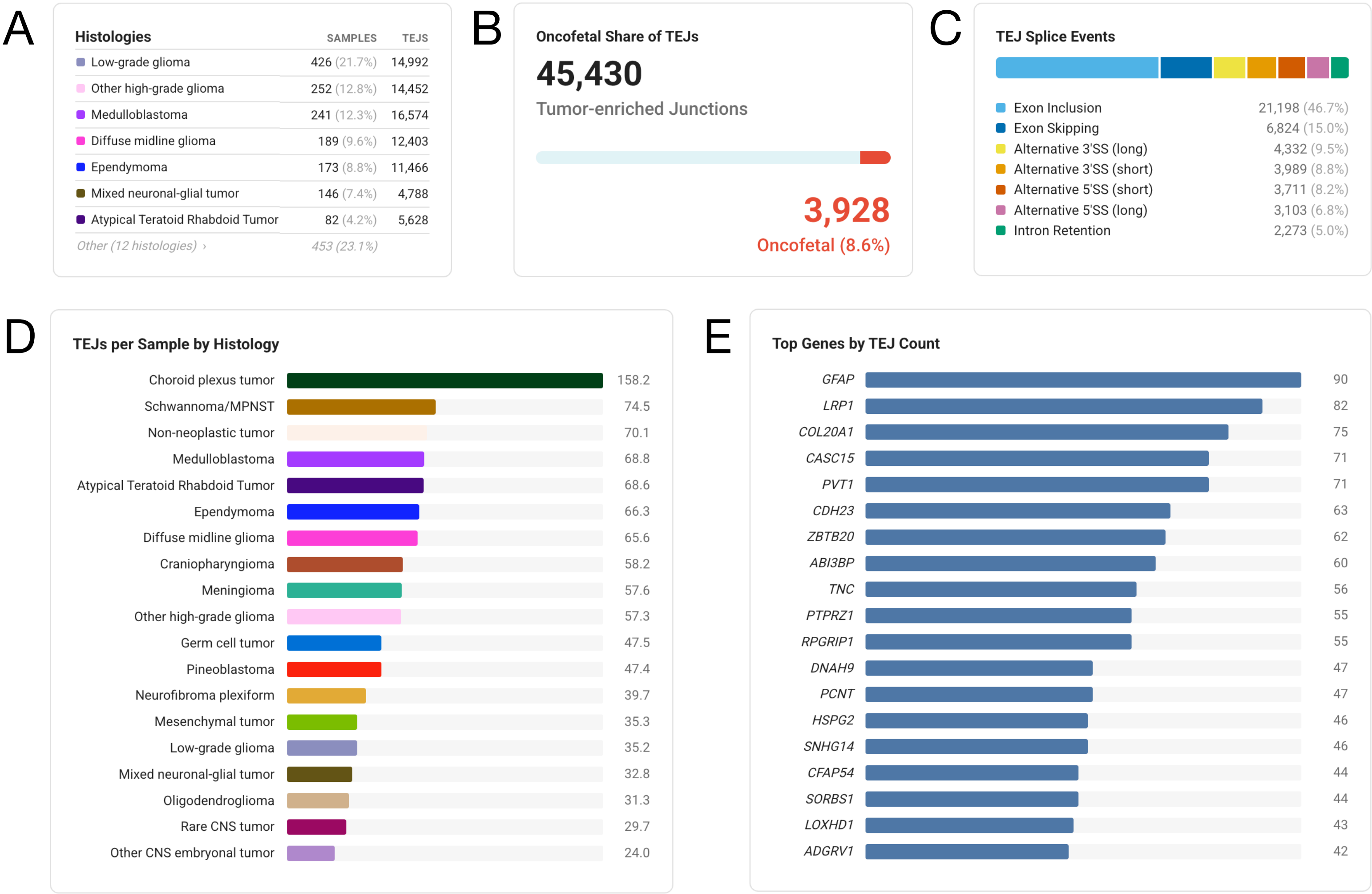
A
Histologies
SAMPLES
TEJS
Low-grade glioma 426 (21.7%) 14,992
Other high-grade glioma 252 (12.8%) 14,452
Medulloblastoma 241 (12.3%) 16,574
Diffuse midline glioma 189 (9.6%) 12,403
Ependymoma 173 (8.8%) 11,466
Mixed neuronal-glial tumor 146 (7.4%) 4,788
Atypical Teratoid Rhabdoid Tumor 82 (4.2%) 5,628
Other (12 histologies) › 453 (23.1%)
B
Oncofetal Share of TEJs
45,430
Tumor-enriched Junctions
3,928
Oncofetal (8.6%)
C
TEJ Splice Events
Exon Inclusion 21,198 (46.7%)
Exon Skipping 6,824 (15.0%)
Alternative 3'SS (long) 4,332 (9.5%)
Alternative 3'SS (short) 3,989 (8.8%)
Alternative 5'SS (short) 3,711 (8.2%)
Alternative 5'SS (long) 3,103 (6.8%)
Intron Retention 2,273 (5.0%)
D
TEJs per Sample by Histology
Choroid plexus tumor 158.2
Schwannoma/MPNST 74.5
Non-neoplastic tumor 70.1
Medulloblastoma 68.8
Atypical Teratoid Rhabdoid Tumor 68.6
Ependymoma 66.3
Diffuse midline glioma 65.6
Craniopharyngioma 58.2
Meningioma 57.6
Other high-grade glioma 57.3
Germ cell tumor 47.5
Pineoblastoma 47.4
Neurofibroma plexiform 39.7
Mesenchymal tumor 35.3
Low-grade glioma 35.2
Mixed neuronal-glial tumor 32.8
Oligodendroglioma 31.3
Rare CNS tumor 29.7
Other CNS embryonal tumor 24.0
E
Top Genes by TEJ Count
GFAP 90
LRP1 82
COL20A1 75
CASC15 71
PVT1 71
CDH23 63
ZBTB20 62
ABI3BP 60
TNC 56
PTPRZ1 55
RPGRIP1 55
DNAH9 47
PCNT 47
HSPG2 46
SNHG14 46
CFAP54 44
SORBS1 44
LOXHD1 43
ADGRV1 42

# Figure 3

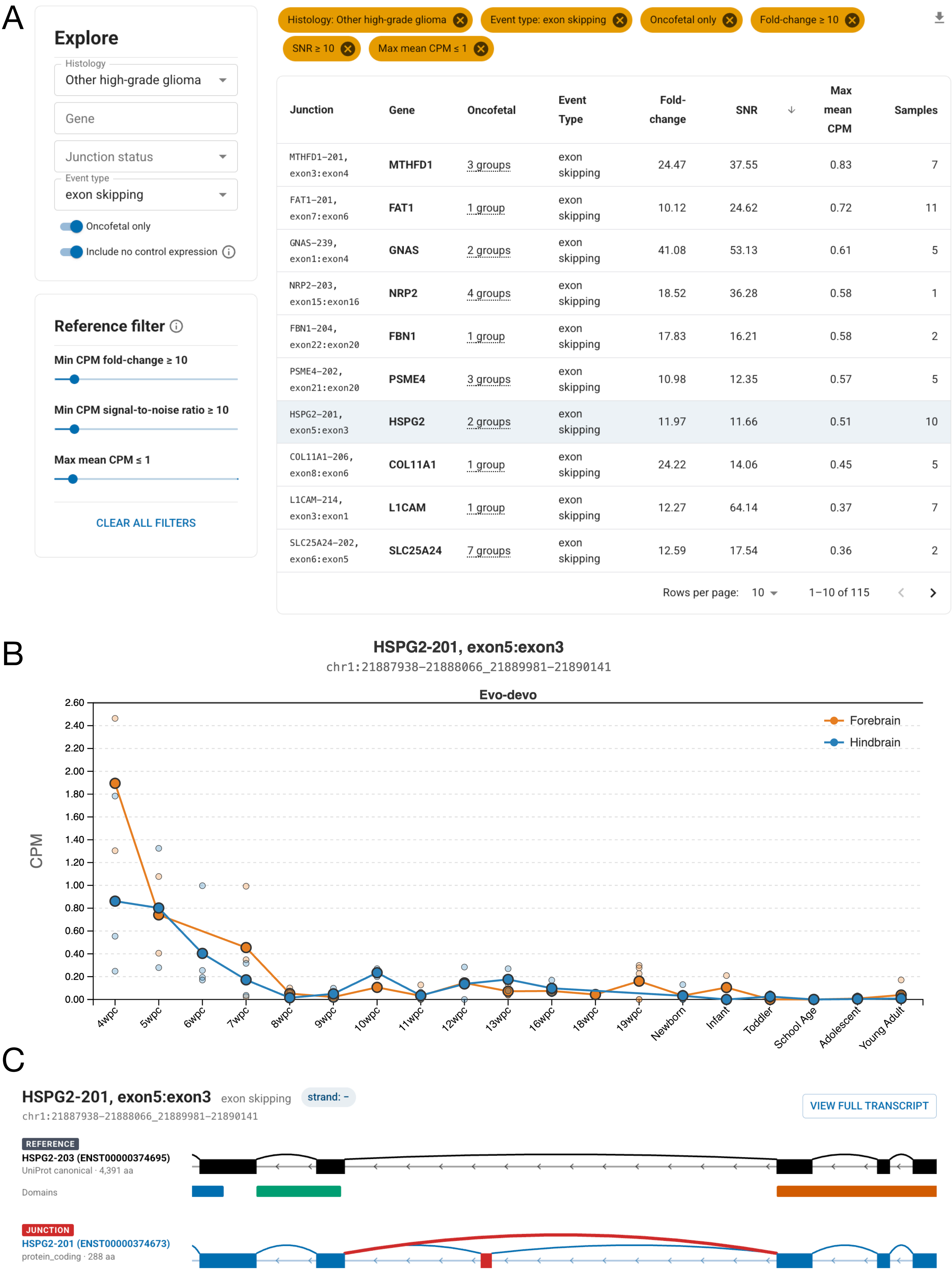

| Junction | Gene | Oncofetal | Event Type | Fold-change | SNR | Max mean CPM | Samples |
|---|---|---|---|---|---|---|---|
| MTHFD1-201, exon3:exon4 | MTHFD1 | 3 groups | exon skipping | 24.47 | 37.55 | 0.83 | 7 |
| FAT1-201, exon7:exon6 | FAT1 | 1 group | exon skipping | 10.12 | 24.62 | 0.72 | 11 |
| GNAS-239, exon1:exon4 | GNAS | 2 groups | exon skipping | 41.08 | 53.13 | 0.61 | 5 |
| NRP2-203, exon15:exon16 | NRP2 | 4 groups | exon skipping | 18.52 | 36.28 | 0.58 | 1 |
| FBN1-204, exon22:exon20 | FBN1 | 1 group | exon skipping | 17.83 | 16.21 | 0.58 | 2 |
| PSME4-202, exon21:exon20 | PSME4 | 3 groups | exon skipping | 10.98 | 12.35 | 0.57 | 5 |
| HSPG2-201, exon5:exon3 | HSPG2 | 2 groups | exon skipping | 11.97 | 11.66 | 0.51 | 10 |
| COL11A1-206, exon8:exon6 | COL11A1 | 1 group | exon skipping | 24.22 | 14.06 | 0.45 | 5 |
| L1CAM-214, exon3:exon1 | L1CAM | 1 group | exon skipping | 12.27 | 64.14 | 0.37 | 7 |
| SLC25A24-202, exon6:exon5 | SLC25A24 | 7 groups | exon skipping | 12.59 | 17.54 | 0.36 | 2 |

# Figure 4

A

| Junction | Gene | Oncofetal | Event Type | Fold-change | SNR | Max mean CPM | Samples |
|---|---|---|---|---|---|---|---|
| NRCAM-202, exon6:exon4 | NRCAM | — | exon skipping | 6.04 | 5.87 | 1.82 | 191 |
| NRCAM-202, exon19:exon17 | NRCAM | — | exon skipping | 6.00 | 11.64 | 3.10 | 144 |

B

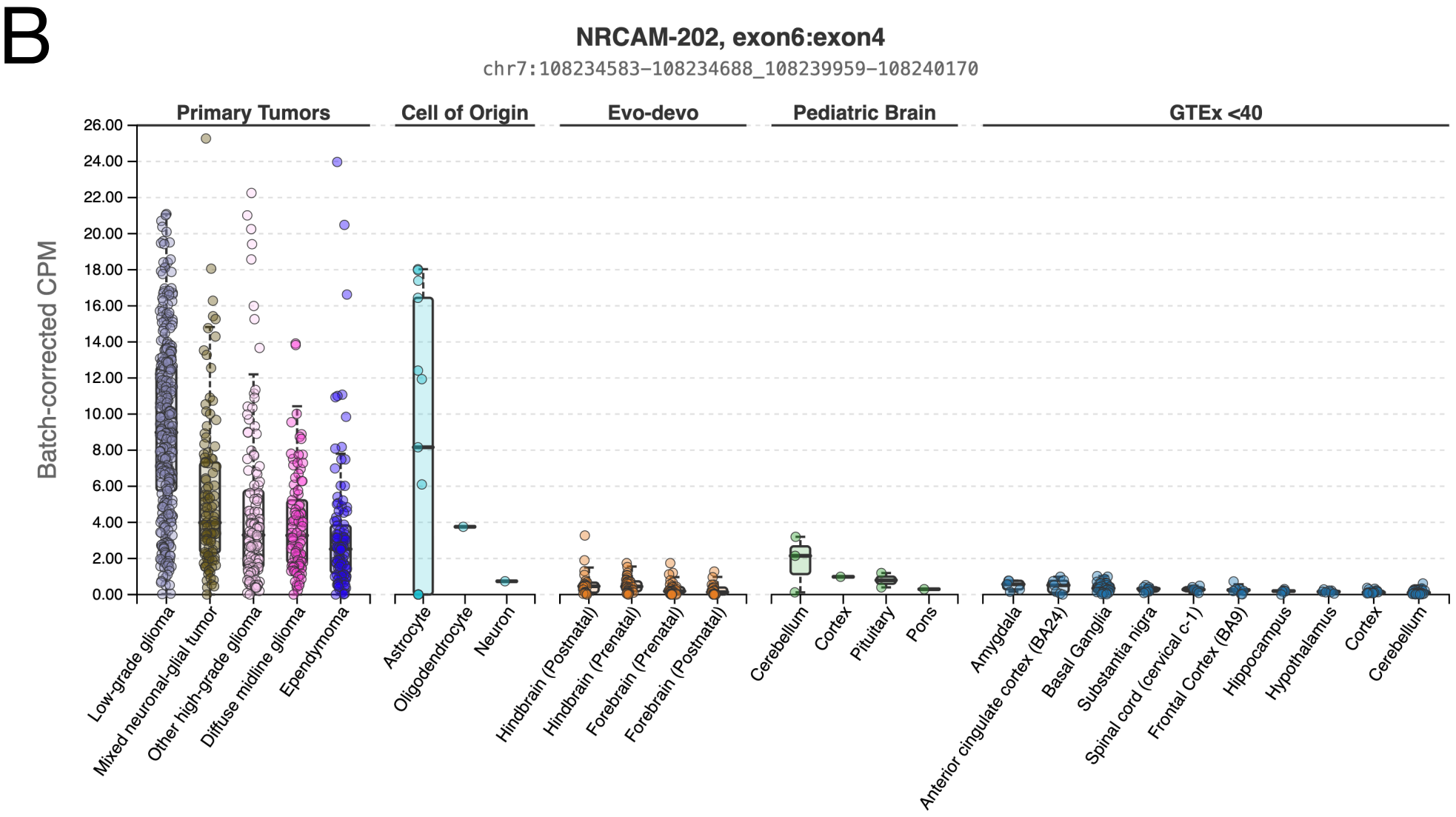

C

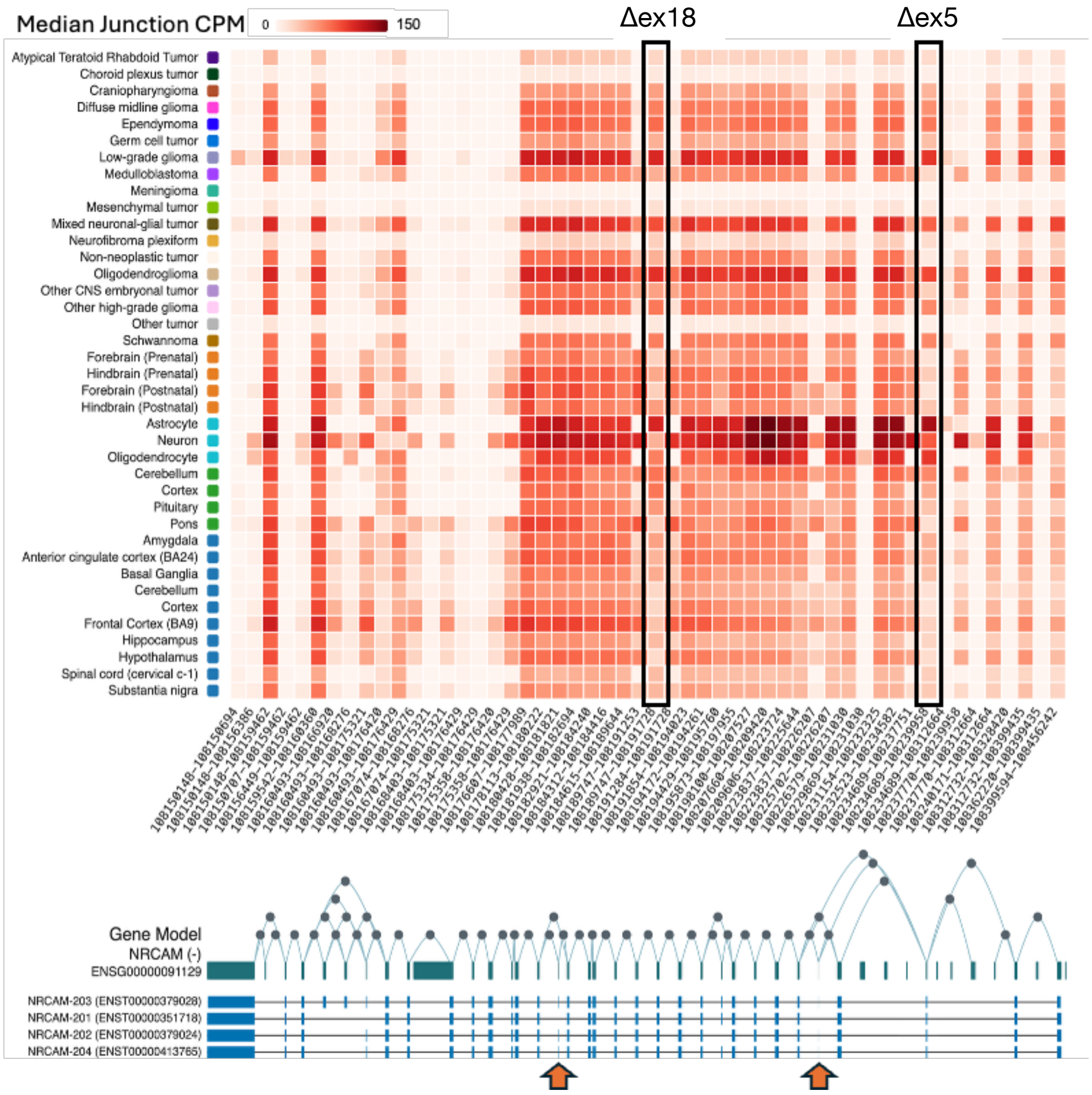